\documentclass[twocolumn,twocolappendix]{aastex701}

\usepackage{booktabs}
\usepackage{amsmath}

\newcommand\kms{km s$^{-1}$}
\newcommand\msun{M$_\odot$}
\newcommand\gizmo{GIZMO}
\newcommand\dice{DICE}
\newcommand\hi{\ion{H}{1}}
\newcommand\hii{\ion{H}{2}}
\newcommand\ovi{\ion{O}{6}}

\newcommand{\rev}[2]{#2}
\newcommand{\revv}[2]{#2}

\begin{document}

\title{\revv{}{The LMC Corona Favors a First Passage}}

\author[0000-0001-9982-0241]{Scott Lucchini}
\affiliation{Center for Astrophysics $|$ Harvard \& Smithsonian, 60 Garden Street, Cambridge, MA 02138, USA}
\email{scott.lucchini@cfa.harvard.edu}

\correspondingauthor{Scott Lucchini}
\email{scott.lucchini@cfa.harvard.edu}

\author[0000-0002-6800-5778]{Jiwon Jesse Han}
\affiliation{Center for Astrophysics $|$ Harvard \& Smithsonian, 60 Garden Street, Cambridge, MA 02138, USA}
\email{jesse.han@cfa.harvard.edu}

\author[0000-0002-4157-5164]{Sapna Mishra}
\affiliation{Space Telescope Science Institute, 3700 San Martin Drive, Baltimore, MD 21218, USA}
\email{smishra@stsci.edu}

\author[0000-0003-0724-4115]{Andrew J. Fox}
\affiliation{AURA for ESA, Space Telescope Science Institute, 3700 San Martin Drive, Baltimore, MD 21218, USA}
\email{afox@stsci.edu}

\begin{abstract}

We use constrained idealized simulations of the LMC/Milky Way interaction to determine if the size of the LMC’s gaseous halo (Corona) can be used to distinguish between first and second passage models
$-$ an orbital trajectory for the LMC in which it has just recently approached the Milky Way for the first time (first passage), or one in which it has had a previous pericenter (second passage). Using live circumgalactic gas particles combined with analytic dark matter potentials evolved to follow previously published orbital trajectories, we find that the first passage model is able to reproduce the observed velocity profile and column density profile of the present day LMC Corona. On the other hand, in a second passage scenario the longer interaction time leads to the velocities and column densities around the LMC at the present day being \rev{too low}{significantly lower than observations}. Based on this observed velocity profile, recent works have found that the LMC's Corona has been truncated to 17$-$20~kpc, and we find truncation radii of $16.6\pm 0.5$~kpc and $5.7^{+1.8}_{-2.2}$~kpc for the first and second passage models, respectively. Thus, based on the gas properties of the LMC's CGM at the present day, a second passage trajectory is \rev{}{strongly} disfavored.

\end{abstract}

\keywords{\uat{Large Magellanic Cloud}{903} --- \uat{Galactic and extragalactic astronomy}{563} --- \uat{Galaxy dynamics}{591} --- \uat{Galaxy physics}{612} --- \uat{Magellanic Clouds}{990} --- \uat{Magellanic Stream}{991} --- \uat{the Milky Way}{1054}}

\section{Introduction} \label{sec:intro}

The Large and Small Magellanic Clouds (LMC, SMC) are the Milky Way's most massive satellites and have the potential to dramatically shape the future evolution of our Galaxy. However, despite significant effort invested in studying them, there are still many unanswered questions regarding their history, which will directly affect their future. The biggest mystery that has persisted for more than 50 years is the question of the period of the LMC's orbit around the Milky Way (MW).

\begin{deluxetable*}{lccccc}
\tablecaption{Galaxy initial conditions}
\label{tab:ics}
\tablehead{ & \multicolumn{3}{c}{DM} & \multicolumn{2}{c}{CGM} \\\cmidrule(lr){2-4} \cmidrule(lr){5-6}
            \colhead{Galaxy} & \colhead{$M_\mathrm{tot}$} & \colhead{$a$} & \colhead{$r_{200}$} & \colhead{$M_\mathrm{tot}$} & \colhead{$M_{r<r_{200}}(t=t_0)$} \\
                & (\msun) & (kpc) & (kpc) & ($10^{9}$~\msun) & ($10^{9}$~\msun)}

 \startdata
MW & $1.1\times10^{12}$ & 22.0 & 199 & 22.2 & 14.9 \\
LMC1 & $1.75\times10^{11}$ & 9.5 & 109 & 5.3 & 1.4 \\
LMC2 & $3.4\times10^{11}$ & 26.3 & 127 & 10.2 & 3.5
\enddata

\tablecomments{LMC1 is used for the first passage orbital model and LMC2 is used for the second passage orbit. Columns (1)$-$(3) provide the inital properties of the dark matter halos of the galaxies. For the LMC2 model, the total mass and scale length change with time (see text). Column (5), $M_{r<r_{200}}(t=t_0)$, lists the amount of CGM material that is within the virial radius of the galaxy when we start the full galaxy interaction simulations. \rev{}{$t_0=5$~Gyr for all models.}}
\end{deluxetable*}

\begin{deluxetable*}{lcccc}
\tablecaption{\rev{}{Orbit Present-day LMC Properties}}
\label{tab:orbits}
\tablehead{ \colhead{Name} & \colhead{(R.A., Decl.)} & \colhead{Distance} & \colhead{PM} & \colhead{RV} \\ & (deg) & (kpc) & (mas yr$^{-1}$) & (km s$^{-1}$)}

\startdata
Gaia$^a$ & $(81.28,-69.78)^b$ & 49.5$^c$ & $(-1.871\pm0.01,0.391\pm0.01)$ & $262.2\pm3.4^d$ \\
K13$^e$ & $(78.76\pm0.52,-69.19\pm0.25)$ & $50.1\pm2.3^f$ & $(-1.91\pm0.02,0.23\pm0.05)$ & $262.2\pm3.4^d$ \\
First Passage$^g$ & $(67.97,-71.68)$ & $52.3$ & $(-2.18\pm0.25,0.20\pm0.29)$ & $262.2\pm18.7$ \\
Second Passage$^h$ & $(75.63,-69.63)$ & $52.0$ & $(-1.81,0.20)$ & $251.5$
\enddata

\tablecomments{$^a$ \citet{gaia21} \revv{}{including systematic uncertainties}\\
$^b$ On sky position used from \citet{vandermarel01}.\\
$^c$ Distance from \citet{pietrzynski19}.\\
$^d$ Radial velocities are taken from \citet{vandermarel02}\\
$^e$ \citet{kallivayalil13}\\
$^f$ Distance from \citet{freedman01}\\
$^g$ \citet{lucchini21}. Uncertainties are estimated based on solar velocity and resolution effects.\\
$^h$ \citet{vasiliev24}}
\end{deluxetable*}

\rev{}{
From decades of detailed observations and sophisticated modeling, we do understand a significant amount about the history of the Magellanic Clouds \citep{donghia16,lucchini24b}. From observations of \hi, we can see the Trailing Stream $-$ a 200$^\circ$-long tail of gas stripped out of the disks of the Clouds \citep{nidever08}. We can also see that the galaxies themselves are in a state of disequilibrium with the LMC's \hi\ disk truncated on the leading edge \citep{salem15} and the main body of the SMC spanning $\sim20$~kpc along the line of sight \citep{murray24,rathore25}. Additionally, the Magellanic Bridge consisting of gas and stars spanning the distance between the galaxies indicates a recent collision \citep{schmidt20} which is corroborated by the SMC's proper motion \citep{zivick18,choi22} and linked star formation bursts several hundred million years ago \citep{harris09,cohen24}.
}

\rev{}{
From absorption spectroscopy towards background quasars, we have also been able to assess the composition and ionization state of the Magellanic System. The bulk of the material has low metallicity ($\sim0.1$~Z$_\odot$) consistent with it coming from the SMC \citep{fox13}, however there is evidence for more enriched material closer to the Clouds indicating that there is some LMC material in the Stream as well \citep{richter13}. Furthermore, an immense amount of ionized material has been detected comoving with the Stream with average ionization fractions of 78\% across the 56 sightlines \citep{fox14}.
}

\rev{}{
In the modern era, there are two paradigms of Stream formation that can explain these observations: the ram-pressure scenario and the tidal scenario. In the ram-pressure scenario, a low-mass LMC approaches the MW with the SMC in an approximately parallel orbit and the ambient CGM of the MW pushes out the LMC and SMC ISM into the Trailing Stream \citep{hammer15,wang19}. However, in order for the sufficient material to be stripped, the LMC's total mass must be quite low ($<2\times10^{10}$~\msun; \citealt{wang22}). At this mass, the LMC is unable to reproduce the MW's reflex motion and density asymmetries \citep{conroy21,garavito-camargo21,petersen21}. In the tidal scenario, the LMC and SMC have been interacting as a binary pair for several billion years and during this time, the tidal interactions have stripped material out of their disks and into the Trailing Stream \citep{besla12,pardy18}. Until recently, these models had difficulty reproducing the observed Stream masses and morphologies. However, with the inclusion of circumgalactic gas around the MW and LMC, the total neutral and ionized stripped material is reproduced \citep{lucchini20}, and we find a turbulent, filamentary morphology as in the observations due to the enhanced hydrodynamic interactions \citep{lucchini21,lucchini24}.
}

\rev{}{
Throughout this paper, we will be exploring the evolution of the Magellanic System within the context of the tidal Stream formation paradigm. However, we would like to highlight a subtlety here: the results presented in this paper are due to ram-pressure interactions, but these ram-pressure interactions are not forming the Stream. We are investigating the present-day properties of the ionized gas in the Magellanic System which originated as the circumgalactic gas around the LMC (the Magellanic Corona) and has been warped and reshaped through ram-pressure forces of the MW CGM acting on it. However, in order for the LMC to have its own CGM it needs to be massive enough for form its own gaseous halo cosmologically ($\gtrsim10^{11}$~\msun; \citealt{jahn22,chisholm25}). This is inconsistent with the ram-pressure Stream formation paradigm. Thus, while this paper is analyzing the effects of ram pressure on the LMC, it is within the context of the Tidal Stream formation paradigm.
}

\rev{}{
This brings us to the question of the period of the LMC's orbit around the MW. Since observations of the Clouds' proper motions revealed surprisingly high tangential velocities \citep{kallivayalil06,kallivayalil13}, it has been assumed that they must be in their first passage around our Galaxy \citep{besla07,besla10,besla12}. 
}
However, recently \citet{vasiliev24} used a genetic algorithm to find an orbital model consistent with the present-day positions and velocities of the LMC while also including an earlier pericentric passage of the LMC around the MW. This has reignited the discussion of the orbital period of the LMC around our Galaxy. The earliest ``many passage'' models remain inconsistent with observations (orbital periods $\lesssim 2$~Gyr; \citealt{mathewson74,gardiner96,yoshizawa03}), but whether the LMC is on its first or second passage around the MW remains unclear.

Furthermore, both first and second passage orbits seem to be consistent with the LMC's orbital constraints from its ejected hypervelocity stars \citep{han25,lucchini25}. About half of the hypervelocity stars in the MW's stellar halo can actually be traced back to the LMC, not the Galactic center \citep{han25}. So in addition to indicating that the LMC harbors its own supermassive black hole, we can use the locations of these stars to trace the LMC's position in the past \citep{lucchini25}. \citet{lucchini25} find that both first passage \citep{lucchini21} and second passage \citep{vasiliev24} trajectories are consistent with the hypervelocity star ejections, however this technique is only reliable back to $\sim800$~Myr ago, and does not account for gas dynamics.

In recent years, we have uncovered many interesting properties of the gas in and around the Clouds \citep{nidever10,fox14,westmeier18,dk22,mishra24}. In addition to the neutral \hi\ visible in radio maps of the southern sky, absorption spectroscopy has revealed an immense amount of ionized material comoving with the Stream. Models and observations have found that this ionized material most likely originated from the LMC's circumgalactic medium (CGM), or Magellanic Corona, that has been stripped and warped through interactions with the MW's own hot CGM \citep{lucchini20,lucchini24,dk22,mishra24}. These ram pressure interactions occur on relatively short timescales and different orbital periods for the LMC should leave unique imprints in the properties of the LMC Corona at the present day.

In particular, \citet{mishra24} found that for all ions studied (\ion{Si}{2}, \ion{Si}{3}, \ion{Si}{4}, and \ion{C}{4}), there was a break in the line of sight (LOS) velocities of the UV absorbers as a function of the impact parameters ($\rho$).
For sightlines with impact parameters $\rho<17$~kpc, the velocities are mostly consistent with the LMC's systemic velocity ($v_\mathrm{LMC}\pm50$~\kms, where $v_\mathrm{LMC} = 280$~\kms\ in their study). However, those sightlines with $\rho>20$~kpc had velocities less than $v_\mathrm{LMC}-50\ \text{km s}^{-1}=230$~\kms, i.e. they were transitioning into the Stream. We define this as the ``truncation radius'' ($\rho_T=17-20$~kpc), the impact parameter at which the line of sight velocities of the LMC CGM drop below 230~\kms. This truncation radius should be quite sensitive to the interactions between the LMC and MW circumgalactic media.

In this work, we perform new hydrodynamic simulations of the first and second passage models including circumgalactic gas around the MW and LMC in order to compare against recent observations of the properties of the LMC's CGM at the present day. In Section~\ref{sec:methods} we describe the simulations and the specific techniques used to constrain the orbital trajectories of the galaxies while self-consistently evolving the live gaseous halos. Section~\ref{sec:results} contains our main results, and we conclude in Section~\ref{sec:disc}.

\begin{figure}
    \centering
    \includegraphics[width=1.0\linewidth]{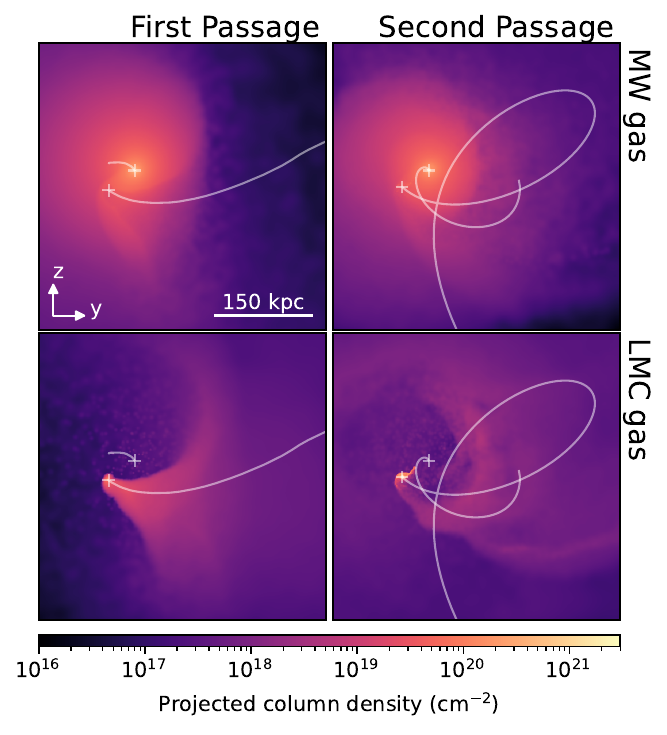}
    \caption{Projected gas density in the first and second passage models at the present day. The top and bottom panels show the MW and LMC CGM gas column density in the Cartesian $y-z$ plane, respectively, with the first passage model on the left and the second passage model on the right. The orbital trajectories of the LMC and MW are drawn in white lines while their present-day positions are marked with plus symbols.}
    \label{fig:cartesian}
\end{figure}

\begin{figure*}
    \centering
    \includegraphics[width=1.0\linewidth]{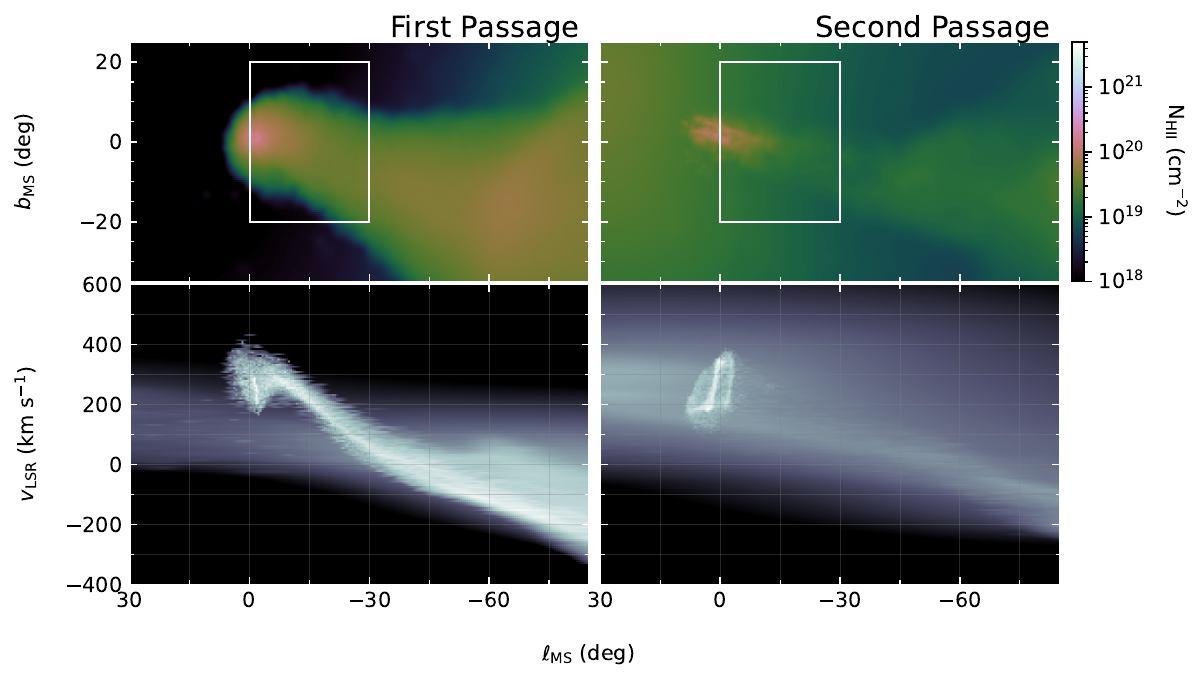}
    \caption{On-sky projection of the present-day Magellanic Corona and its velocity profile. The top panels show the \hii\ gas column density in Magellanic Coordinates for the first passage model on the left and the second passage model on the right. The white box denotes the region from which the random sightlines were selected to best match the region probed by the observational data in \citet{mishra24}. The bottom panels show the gas LSR velocity distribution as a function of Magellanic Longitude for the first (left) and second (right) passage models.}
    \label{fig:magellanic}
\end{figure*}

\section{Methods} \label{sec:methods}

These simulations were run using \gizmo, employing its ``meshless finite-mass'' (MFM) hydrodynamics solver which aims to mitigate some of the issues with smoothed particle hydrodynamics (SPH) while maintaining its Lagrangian nature \citep{hopkins15,springel05}. \gizmo\ was compiled as done in \citet{lucchini24} using adaptive gravitational softenings for gas, star formation, mechanical stellar feedback, and radiative cooling down to low temperatures via metal lines. Stars are formed out of gas cells following \citet{sh03} above a density threshold of 100~cm$^{-3}$, assuming it is self-gravitating \citep{hopkins13} and converging ($\nabla\cdot\vec{v}<0$). Supernovae return mass, energy, and momentum to their surroundings through direct mechanical feedback \citep{hopkins14,hopkins18} with parameters following the setup used in the AGORA project \citep{agora} $-$ a constant supernova rate of $3\times10^{-4}$~SNe~Myr$^{-1}$~M$_\odot^{-1}$ for all stars younger than 30~Myr old, injecting 14.8~\msun\ with $10^{51}$~erg of energy and metals. Radiative cooling follows \citet{hopkins18-fire} including metal lines and fine-structure and molecular cooling down to 10~K \citep{wiersma09,hopkins18-fire,hopkins23}.

Additionally, we have added the ability to include analytic potentials that follow prescribed orbital trajectories. These analytic potentials follow Hernquist profiles \citep{hernquist90} and can account for time dependent masses and scale lengths. In this work, we use the orbits from \citet{lucchini21} and \citet{vasiliev24} which we designate ``first passage'' and ``second passage'' respectively. On top of these analytic potentials, we allow the gas to evolve self-consistently following the MFM method in \gizmo. We describe these new simulations as ``constrained idealized simulations'' because the orbital trajectories are fixed while the hydrodynamics is live.
The trajectories of the orbits used in this work (along with the gas densities at the present day) are shown in Figure~\ref{fig:cartesian} as the white lines. The plus marks denote the galaxies' positions at the present day.

Table~\ref{tab:ics} lists the initial propererties of the simulated galaxies used in this work.
In the first passage orbit, the mass and scale lengths of the galaxies remain fixed at $M_\mathrm{MW}=1.1\times10^{12}$~\msun, $a_\mathrm{MW}=22$~kpc, $M_\mathrm{LMC}=1.75\times10^{11}$~\msun, and $a_\mathrm{LMC}=9.5$~kpc (LMC1). In the second passage orbit, the MW again remains fixed with $M_\mathrm{MW}=1.1\times10^{12}$~\msun\ and $a_\mathrm{MW}=22$. However, the LMC's mass follows the amount of bound material as defined in the \citet{vasiliev24} simulations. At each timestep, we numerically fit the radial profile of the bound $N$-body particles with a Hernquist profile via a Trust Region Reflective least-squares routine implemented in Python's \texttt{scipy.curve\_fit} function. This gives us an initial LMC with $M_\mathrm{LMC}=3.4\times10^{11}$~\msun\ and $a_\mathrm{LMC}=26.3$~kpc. The mass decreases similarly to what is shown in the bottom panel of figure~3 in \citet{vasiliev24}, resulting in a present-day LMC with $M_\mathrm{LMC}=1.0\times10^{11}$~\msun\ and $a_\mathrm{LMC}=7.9$~kpc.

On top of these analytic potentials, we include live gas particles representing a gaseous disk and halo. The MW models are the same between the two orbits, so we use the same initial MW CGMs which were originally used in \citet{lucchini24}. It is initialized following a beta profile ($\rho\propto\left( 
1+(r/r_c)^2 \right)^{-3\beta/2}$; \citealt{salem15}) with $r_c=0.35$ and $\beta=0.559$. It starts with a total mass of $2.2\times10^{10}$~\msun\ \rev{}{(2\% of the galaxy's DM mass)} at $10^6$~K. After \rev{4}{5}~Gyr in isolation (with the analytic potential described above), $1.5\times10^{10}$~\msun\ remains within $r_{200}=199$~kpc ($1.9\times10^{10}$~\msun\ remains bound), and the CGM has a mean temperature of $5.1\times10^6$~K. Continuing the evolution in isolation after this point, the total gas mass within $r_{200}$ changes by $<4\%$ over the subsequent 4~Gyr. Thus, we use $t_0=5$~Gyr as the initial snapshot for our interacting simulations.

\rev{}{The LMC initial conditions are built using the \dice\ code \citep{perret14}\footnote{\url{https://bitbucket.org/vperret/dice/src/master/}}. We include a Hernquist DM potential in the creation of the initial conditions (ICs), and then excise the DM particles before running the simulation.}
We use two different initial Coronae models due to the difference in initial mass between the two models. For the first passage model, we again follow the setup from \citet{lucchini24} using an isothermal profile with a mass of $5.3\times10^9$~\msun\ \rev{}{(3\% of the galaxy's DM mass)} and temperature of $5\times10^5$~K. After \rev{6}{5}~Gyr in isolation (again using the analytic potential), $1.4\times10^{9}$~\msun\ remains within $r_{200}=109$~kpc (with $3.2\times10^9$~\msun\ bound), and the CGM has a median temperature of $1.0\times10^6$~K.
For the second passage model, we again initialize the LMC with a Hernquist DM profile and an isothermal CGM. We begin with a $1.0\times10^{10}$~\msun\ CGM at $5\times10^5$~K. This is \rev{}{again 3\% of the DM mass of the galaxy which is }increased from our first passage model\rev{ due to the higher initial mass of the LMC in this model}{} ($3.4\times10^{11}$ vs $1.8\times10^{11}$~\msun). After \rev{6}{5}~Gyr, $3.5\times10^9$~\msun\ remains within $r_{200}=127$~kpc (with $7.9\times10^9$~\msun\ bound), with a median temperature of $1.3\times10^6$~K. For both of these LMC models, continued evolution in isolation for \rev{4}{5} more Gyr results in the total gas mass within $r_{200}$ changing by $<3\%$.

\begin{figure*}
    \centering
    \includegraphics[width=0.875\linewidth]{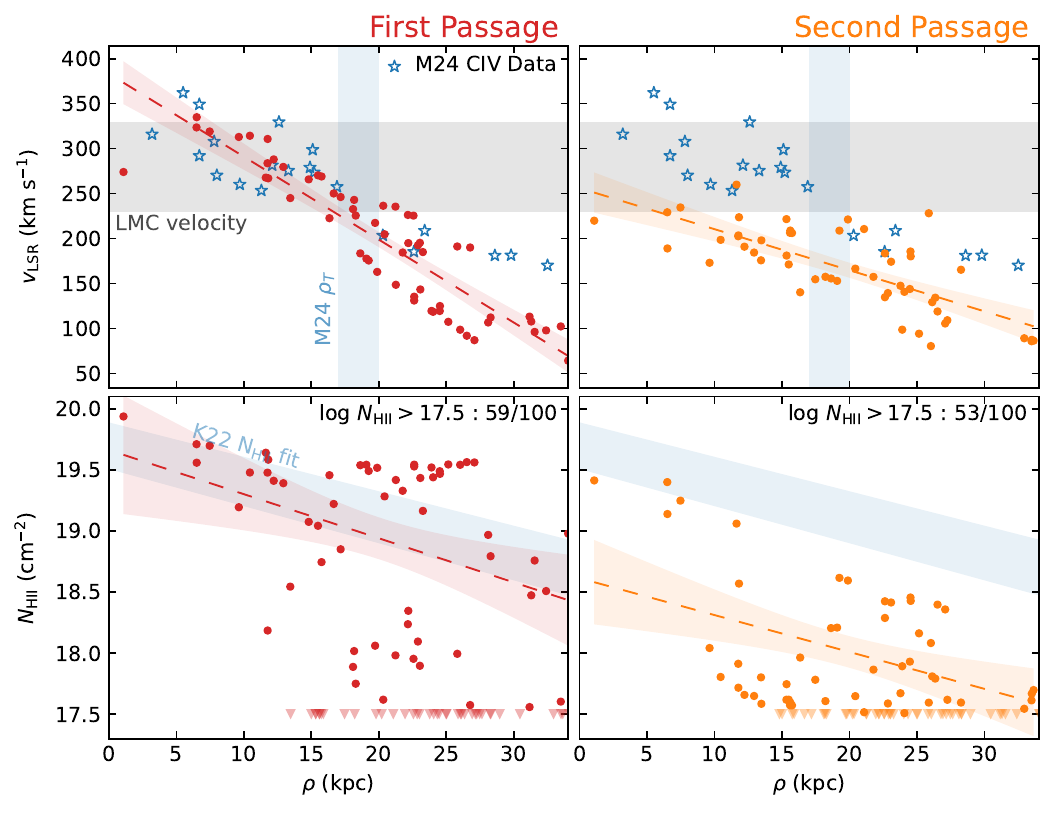}
    \caption{LSR velocities and column densities of mock observations of the simulations compared against the data. The top panels show the column density weighted LSR velocities, and the bottom panels show the \hii\ column densities, both as a function of impact parameter from the LMC. The left panels show the results for the first passage model (in red), while the right panels are for the second passage model (in orange). The best fit linear regression to the data points is shown as a dashed line with the 95\% confidence interval shown as the red/orange shaded region. The blue stars in the top panels are the observational data points from \citet{mishra24} for the \ion{C}{4} detections and the blue vertical band is their quoted truncation radius. The grey horizontal band shows the LMC's systemic velocity $\pm50$~\kms\ which was used to determine the truncation radius (see text). In the bottom panels, the blue region is the fit from \citet{dk22} after using \textsc{cloudy} modeling to extrapolate the \hii\ densities. For the top panels and both fits, we have only included mock sightlines with \hii\ column densities above $10^{17.5}$~cm$^{-2}$. The downward arrows in the bottom panels show the $\rho$ values for the sightlines with log~$N_\mathrm{HII}<17.5$ which were not used in the analysis. There were 51 and 48 sightlines remaining for the first and second passage models, respectively.}
    \label{fig:velfits}
\end{figure*}

\subsection{Mock observations}

In order to compare against the data in \citet{mishra24}, we perform mock spectroscopic observations through our simulations. For this we use the \textit{Trident} code\footnote{\url{https://github.com/trident-project/trident}} \citep{hummels17}, which is built upon \textit{yt}\footnote{\url{https://yt-project.org}} \citep{turk11}. \textit{Trident} uses pre-computed \textsc{cloudy} tables to populate the simulation with mass fractions of a large variety of ions \citep{ferland13}. Using these mass fractions, it then can compute the column densities and projected gas velocities along lines of sight through the simulation.

We randomly select 100 sightlines originating at the solar location and extending towards a direction with Magellanic Longitude between $-30$ and 0 degrees, and Magellanic Latitude between $-30$ and 30 degrees (using Magellanic Coordinates as defined in \citealt{nidever08}; this region is shown as a white box in Figure~\ref{fig:magellanic}). This region is selected to match up with the observational area probed by the background quasars analyzed in \citet{mishra24}. We then calculate the total \hii\ column density and the column density weighted velocity in the local standard of rest (LSR) frame. We keep values from all sightlines with total \hii\ columns greater than $10^{17.5}$~cm$^{-2}$. This results in 59 data points for the first passage model, and 53 data points for the second passage model.

\section{Results} \label{sec:results}

\begin{figure}
    \centering
    \includegraphics[width=0.8\linewidth]{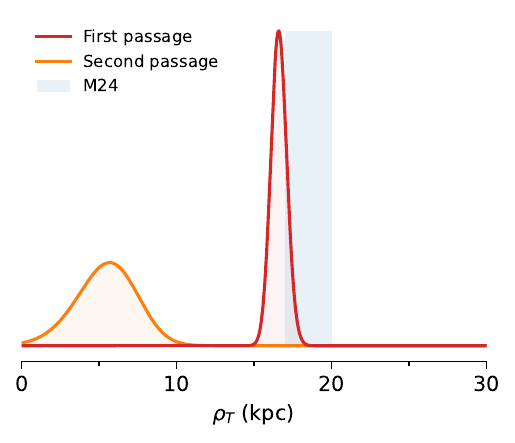}
    \caption{Distributions of truncation radii ($\rho_T$) for the first and second passage models compared against the range found in \citet{mishra24} ($17-20$~kpc, shown in blue). As before, the first passage model is shown in red, and the second passage model is shown in orange. The curves are presented as skewnorm distributions where we have calculated the root semivariances by finding the range of $\rho$ values where the 68\% confidence interval of the $v_\mathrm{LSR}$ vs $\rho$ fit crosses the $v_\mathrm{LMC}-50$~\kms\ $=230$~\kms\ line (see top row of Figure~\ref{fig:velfits}). We find $\rho_T=16.6\pm0.5$~kpc for the first passage model, and $\rho_T=5.7^{+1.8}_{-2.2}$~kpc for the second passage model.}
    \label{fig:trunc}
\end{figure}

The physical properties of the gas at the present day in our simulations are shown in Figures~\ref{fig:cartesian} and \ref{fig:magellanic}. Figure~\ref{fig:cartesian} shows the $y-z$ cartesian projected column densities for gas originating in the MW's CGM on the top, and the LMC's CGM, or Magellanic Corona, on the bottom. The left panels show the first passage model and the right panels show the second passage model. The white lines show the trajectories of the MW and LMC in the two models with the plus marks showing the present-day positions of the galaxies.

Figure~\ref{fig:magellanic} shows the on-sky projected column densities in Magellanic Coordinates (as defined in \citealt{nidever08}) as well as the gas LSR velocities as a function Magellanic Longitude. Again, the left panels show the results for the first passage model, while the right panels show the results for the second passage model. As stated above, the white box in the top panels shows the region from which the 100 sightlines were randomly selected for the mock observations.

Both of these figures show that after a second passage around the MW, even one at $\sim100$~kpc 6~Gyr ago, the Magellanic Coronal material is much more widely distributed and diffuse. While in the first passage model, we can clearly see the generation of a bow shock \citep{setton23,carr25} and collimation of the Corona in a tail behind the LMC as a result of the ram pressure from the MW's ambient CGM. However, the top panels of Figure~\ref{fig:cartesian} show that the MW's CGM is also not unaffected. The LMC's approach has a dramatic impact on our Galactic atmosphere and its distribution, very similar to the results already detected in the stellar halo \citep{conroy21}.

Figure~\ref{fig:velfits} shows the column density-weighted line-of-sight velocities (top) and total integrated column densities (bottom) for each sightline with $N(\mathrm{H\ \textsc{ii}})>10^{17.5}$~cm$^{-2}$. In the top panels, the blue stars are the results from the \ion{C}{4} observations in \citet{mishra24} with the blue vertical band showing their quoted truncation radius ($\rho_T=17-20$~kpc). The grey horizontal band denotes the LMC's systemic velocity plus or minus 50~\kms\ ($230-330$~\kms).

The bottom panels of Figure~\ref{fig:velfits} shows the column density profiles compared against the $N(\mathrm{H\ \textsc{ii}})$ fit from \citet{dk22} where they used \textsc{cloudy} modeling to estimate the total \hii\ columns. They compute these values and provide linear fits for the cool, photoionized CGM, the warm, interface layers, and the warm-hot, ambient Magellanic Corona. Since we have only included a warm-hot, single-phase CGM in these models, we are comparing against the profile derived from \ovi\ observations.

In our simulations, we find that the velocities and column densities of the LMC's CGM in the first passage scenario are consistent with the observations, while the velocities and column densities in the second passage scenario are systematically lower than the observations. This is shown in the left and right panels of Figure~\ref{fig:velfits}, respectively. We also show the linear fits and 95\% confidence intervals for the two orbits.

In order to quantify this, we look at the likelihood distribution of truncation radii for our two models in Figure~\ref{fig:trunc}. Again $\rho_T$ from \citet{mishra24} is shown in blue, and the two gaussian distributions are derived from the values of $\rho$ where the linear fits and confidence intervals in Figure~\ref{fig:velfits} intersect the lower limit of the LMC's velocity, 230~\kms. Note that Figure~\ref{fig:velfits} shows the 95\% confidence interval, while in determining the skewnorm distributions shown in Figure~\ref{fig:trunc}, we used the 68\% confidence interval.
We find $\rho_T=15.7^{+0.7}_{-1.0}$~kpc for the first passage model, and $\rho_T=8.5\pm1.5$~kpc for the second passage model (where the uncertainties are the 1$\sigma$ root semivariances). This shows that the first passage scenario is much more consistent with the observations than second passage ($1.3\sigma$ vs $5.7\sigma$).

\section{Discussion \& Conclusions} \label{sec:disc}

In this work, we have presented evidence that the LMC is on its first passage around the Milky Way based on the morphology, column density, and velocity of its CGM in absorption at the present day. We have performed constrained idealized simulations with analytic DM potentials explicitly following two different orbital trajectories from the literature: one with the LMC on first passage \citep{lucchini21}, and one on second passage \citep{vasiliev24}. On top of these analytic models, we have allowed the MW and LMC circumgalactic gas to evolve self-consistently following \gizmo's MFM scheme.

At the present day, we find that the second passage model results in the LMC's CGM being too diffuse and widespread on the sky, with LSR velocities that are too low. This is a result of the significantly longer timescales under which the LMC and MW CGMs have been interacting.

As shown in Figure~\ref{fig:magellanic}, the LMC Corona is much more contained and defined in the first passage model. Additionally, a bow shock is visible on the leading side. With new observations on the front side of the LMC we will hopefully be able to determine the offset between the ionized material and the edge of the LMC's disk giving us important constraints on the mass ratio between the MW and LMC CGMs. We have an approved HST/COS Cycle 32 program to probe the LMC CGM in this region (PI: S. Mishra).

\rev{}{
In addition to the bow shock, the leading edge of the LMC's \hi\ disk has been shown to be truncated \citep{salem15}. Prior to the discovery of the Magellanic Corona, this was explained as due to ram-pressure effects of the MW's CGM acting on the LMC's disk ($P\propto\rho v^2$; \citealt{salem15}). In this scenario, the LMC's disk experiences a fast-moving, low-density wind which provides the pressure force. A similar explanation applies within the Magellanic Corona paradigm. In this case, the Magellanic Corona is a mediator of the interactions between the MW CGM and LMC disk. The MW CGM provides the ram-pressure force to reshape the Magellanic Corona (as shown in Figures \ref{fig:cartesian} and \ref{fig:magellanic}, left panels). This compresses the Coronal gas on the leading edge of the Magellanic System increasing its density. The Magellanic Corona then provides a ram-pressure force on the LMC's disk. While the relative velocities between the Corona and the disk are much lower in this scenario, the density of the Corona is much higher than the MW CGM because it has been compressed and reshaped. This could result in an equivalent ram-pressure force being applied to the LMC's \hi\ disk. Due to the simplistic nature of the simulations presented here, we cannot directly explore the morphologies of the LMC's disk, but this will be investigated in future work.
}

\rev{}{
The response of the MW is also more consistent with a first passage scenario. \citet{sheng24} found that a previous LMC pericenter within the past 5~Gyr is inconsistent with the observed reflex motion, however earlier pericenters give the MW enough time to resettle into equilibrium resulting in the same present-day signal \citep{vasiliev24}. Also, while the parameter space has not been fully explored, the MW halo's observed dynamical response \citep{conroy21} agrees well with first passage models \citep{garavito-camargo19,garavito-camargo21}.
}

Recently, \citet{zhu24} explored the effect of ram-pressure stripping on dwarf galaxies in a MW-like environment using wind-tunnel simulations. While the hydrodynamic simulations were restricted to low mass dwarfs, they did extend their calculations up to LMC mass scales analytically and they find truncation radii of $\sim10-15$~kpc depending on the density profile of the LMC's CGM. Fitting our stable LMC CGM with an isothermal profile as used in \citet{zhu24} ($\rho(r)=\rho_0(r/r_0)^\alpha$) gives a value of $\alpha=-2.2$ which would extend the truncation radius out to slightly larger radii ($\rho_T\gtrsim15$~kpc, consistent with our results.

Figures~\ref{fig:velfits} and \ref{fig:trunc} are comparing against the 17$-$20~kpc range of $\rho_T$ quoted in \citet{mishra24}. This value takes into account all the ions they observed, \ion{Si}{2}, \ion{Si}{3}, \ion{Si}{4}, and \ion{C}{4}. However, we can also perform the fitting and confidence interval calculation for the observational data just as we have done with the simulation. For this we have just used the \ion{C}{4} data shown in Figure~\ref{fig:velfits} and we find $\rho_T=20.4^{+1.5}_{-0.9}$, consistent with the high end of their estimate. Thus, we can be confident that our fitting process is accurately tracing the truncation radius.

Throughout this work we have neglected the effects of the SMC on the evolution of the LMC's Corona. While this may be accurate to first order, there could be many additional effects due to the SMC's interactions. In our models presented here, there is no neutral Trailing Stream since it is primarily sourced from material stripped out of the SMC through tidal interactions with the LMC. However, this stripped neutral material will interact with the surrounding LMC and MW circumgalactic gas and could change these density and velocity profiles through mixing.

\rev{}{
Upcoming observations of the Magellanic Corona in \ovi\ emission with the Aspera NASA Pioneer mission will give us a direct look at the morphology and distribution of the hot Magellanic gas. Furthermore, as more and more fast radio bursts (FRBs) are identified, investigating a correlation between dispersion measure within the extent of the Corona and outside could give constraints on the amount of material and its location.
}

\rev{Through the combined use of hypervelocity stars \citep{lucchini25} and the hydrodynamic analysis in this paper,}{With these new constrained idealized simulations of the LMC's interactions with our Galaxy}, we have constrained the LMC's orbital trajectory over the past several billion years. \rev{}{We have shown that a first passage orbital trajectory for the LMC is more consistent with the present-day observed gas dynamics than a second passage scenario.} In future work, we will apply a similar constrained idealized simulation technique to the family of possible SMC orbits to better understand the details of the formation of the neutral Stream.

\begin{acknowledgements}
    S.L. would like to thank Eric Koch and the members of Seamless Astronomy for workshopping the figures in this paper. Support for S.L. was provided by Harvard University through the Institute for Theory and Computation Fellowship. The computations in this paper were run on the FASRC cluster supported by the FAS Division of Science Research Computing Group at Harvard University.
\end{acknowledgements}

\software{\dice\ \citep{perret14}, \gizmo\ \citep{hopkins15,springel05}, matplotlib v3.10.0 \citep{hunter07}, numpy \citep{harris20}, scipy \citep{virtanen20}, \textit{Trident} v1.3 \citep{hummels17}, \textit{yt} v4.5 \citep{turk11}}

\appendix

\section{Parameter Space Exploration}
\rev{}{
To test the robustness of the result, we explored additional initial masses for the MW and LMC CGMs. The fiducial setup above used a CGM masses corresponding to 2\% and 3\% of the total DM masses for the MW and the LMC, respectively. This leads to total initial CGM gas masses of $2.2\times10^{10}$~\msun\ for the MW, $5.3\times10^9$~\msun\ for the first passage LMC, and $1.0\times10^{10}$~\msun for the second passage LMC. In this parameter study, we expanded these values to 0.5\% and 6\% of the DM mass for the MW, and 1\% and 6\% of the DM mass for the LMC. As in our fiducial case, the galaxies stabilize after 5~Gyr in isolation. Table~\ref{tab:appendixmasses} lists the initial and final gas masses for these additional runs. Beyond CGM masses of 6\% the DM halo mass, the CGM was not able to reach a stable state after 10~Gyr of evolution. Lower masses were stable, however, they result in lower star formation rates and gas disk masses.
}

\begin{deluxetable}{lccc}
\tablecaption{\rev{}{Galaxy initial conditions}}
\label{tab:appendixmasses}
\tablehead{\colhead{Galaxy} & \colhead{$f_\mathrm{CGM}$} & \colhead{$M_\mathrm{tot}(t=0)$} & \colhead{$M_{r<r_{200}}(t=t_0)$}\\
            &  & ($10^9$~M$_\odot$) & ($10^9$~M$_\odot$)}

\startdata
MW &  &  &  \\
 & 0.5\% & 5.6 & 3.9 \\
 & 2\% & 22.2 & 14.9 \\
 & 6\% & 66.7 & 39.4 \\
LMC1 &  &  \\
 & 1\% & 1.8 & 0.5 \\
 & 3\% & 5.3 & 1.4 \\
 & 6\% & 10.5 & 2.8 \\
LMC2 &  &  \\
 & 1\% & 3.4 & 1.2 \\
 & 3\% & 10.2 & 3.5 \\
 & 6\% & 20.4 & 5.2
\enddata

\tablecomments{LMC1 is used for the first passage orbital model and LMC2 is used for the second passage orbit. Column (2) lists the ratio of the total CGM gas mass to the DM mass. Columns (3) and (4) present the total gas masses. $t_0=5$~Gyr for all models.}
\end{deluxetable}

\begin{figure*}
    \centering
    \includegraphics[width=0.8\linewidth]{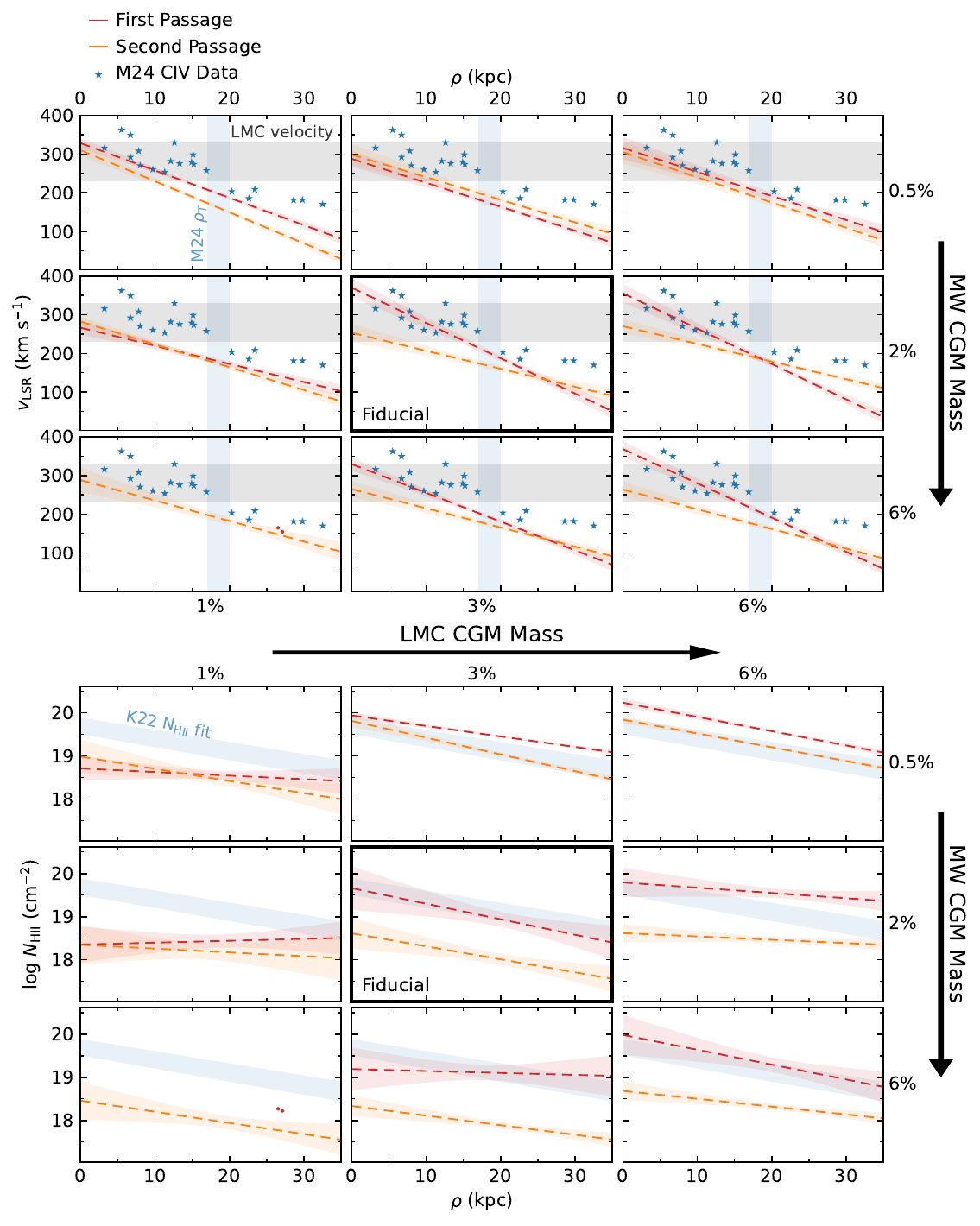}
    \caption{\rev{}{Velocity and column density profiles for the present-day Magellanic Corona gas varying the initial total mass of the MW and LMC CGMs. Results from the first passage orbits are shown in red and second passage is shown in orange. The left panels show the line of sight velocity as a function of impact parameter away from the LMC in comparison with the observational data \citep{mishra24}. Analogous to the top panels in Figure~\ref{fig:velfits}, however we are just showing the fits and 95\% confidence intervals for clarity. The grey horizontal band shows the systemic velocity of the LMC ($\pm50$~\kms), and the blue vertical band shows the truncation radius estimated from the observational data in \citet{mishra24}. The right panels show the \hii\ column densities as a function of impact parameter with the blue band denoting the range of observational values estimated from \citet{dk22}. The center panel shows the results for our fiducial simulation presented in the main article. Note that there is no fit for the first passage simulation with 1\% LMC mass and 6\% MW mass panel (red line, lower left). This is because there are only two sightlines that have column densities above 10$^{17.5}$~cm$^{-2}$, thus a confidence interval can't be calculated. We show the two data points in those panels.}}
    \label{fig:bulk}
\end{figure*}

\rev{}{
We performed the full three-body constrained idealized simulations will all combinations of these initial galaxies. This resulted in nine first passage simulations plus nine second passage simulations. Figure~\ref{fig:bulk} shows the velocity and column density fits for the grid of simulations. The left grid of panels show the line of sight velocity as a function of impact parameter away from the LMC. In the low MW CGM mass case (0.5\%) the velocity profiles of the first and second passage orbits are comparable, but they begin to deviate with higher MW CGM masses. The LMC CGM mass doesn't have a strong effect on the second passage velocity profile$-$it is too low in all cases. However, in the first passage case, higher LMC CGM masses generally have steeper slopes, more consistent with the inner 20~kpc of the observational profile. The right grid of panels shows the column density profile as a function of impact parameter. As expected, here we see higher column densities for more massive LMC Coronae, while the observed columns generally decrease with increasing MW CGM mass due to the increased ram pressure.
}

\rev{}{
The only cases in which the second passage scenario provides a better match to the data are (MW, LMC) CGM mass values of (0.5\%, 3\%) (center top), (2\%, 1\%) (left middle), and (6\%, 1\%) (left bottom). In the (0.5\%, 3\%) case, both trajectories provide poor fits to the velocity profiles, while in the (2\%, 1\%) case, the column density profiles for both trajectories are incorrect. In the (6\%, 1\%) case , there are not enough sightlines with column densities above $10^{17.5}$~cm$^{-2}$ in the first passage trajectory to provide a confidence interval for the fit. Thus this setup also prefers the second passage model, however its column densities are again too low. In all other cases, the truncation radius of the first passage model is closer to the observed estimate of $17-20$~kpc than the second passage model. Furthermore, the fiducial model (center panel) provides the best fit to the column density profile.
}

\begin{figure}
    \centering
    \includegraphics[width=0.42\textwidth]{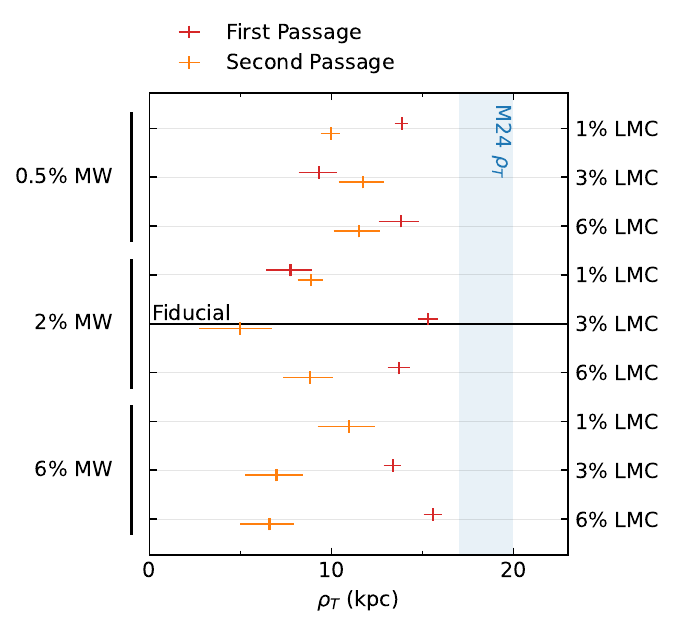}
    \caption{\rev{}{Truncation radii for the different simulations in comparison to the observed value from \citet{mishra24} shown as a blue band. Each row is a different combination of MW and LMC CGMs and the mean and 1$\sigma$ extent for the first and second passage orbits are shown in red and orange, respectively. The fiducial model (2\% MW, 3\% LMC) is highlighted with a black line. As in Figure~\ref{fig:bulk}, the 6\% MW, 1\% LMC first passage simulation doesn't have enough sightlines above 10$^{17.5}$~cm$^{-2}$ in column density and thus an estimate for the truncation radius cannot be made.}}
    \label{fig:bulktrunc}
\end{figure}

\rev{}{
Figure~\ref{fig:bulktrunc} shows the truncation radii of the different models compared against the observations. The truncation radius is defined as the impact parameter value at which the velocity profile crosses $v_\mathrm{LMC}-50$~\kms\ $=230$~\kms. As in Figure~\ref{fig:trunc}, first passage orbits are shown in red and second passage in orange, except instead of displaying a full distribution, each model's $\mu$ and $\sigma$ is depicted as the vertical mark and error bars. The fiducial model is marked with a black line.
}

\bibliography{references}{}
\bibliographystyle{aasjournalv7}

\end{document}